\documentclass[aps,twocolumn,showpacs,preprintnumbers,amsmath,amssymb]{revtex4-1}
\usepackage{amssymb}
\usepackage{graphicx}% Include figure files
\usepackage{dcolumn}% Align table columns on decimal point
\usepackage{bm}
\usepackage{subfigure}

\begin{document}

\title{Giant anomalous Nernst effect in the magnetic Weyl semimetal Co$_3$Sn$_2$S$_2$}

\author{Haiyang Yang$^{1}$, Wei You$^{1}$, Jialu Wang$^{1}$, Junwu Huang$^{1}$, Chuanying Xi$^{3}$, Chao Cao$^{1,*}$, Mingliang Tian$^{3}$, Zhu-An Xu$^{2}$, Jianhui Dai$^{1,*}$, Yuke Li$^{1,*}$}

\affiliation{$^{1}$ Department of Physics and Hangzhou Key Laboratory of Quantum Matters, Hangzhou Normal University, Hangzhou 311121, China\\
  $^{2}$ Department of Physics, Zhejiang University, Hangzhou 310027, China\\
  $^{3}$ High Magnetic Field Laboratory, Chinese Academy of Sciences, Hefei 230031, China
\\}
%\date{}

\begin{abstract}
In ferromagnetic solids, even in absence of magnetic field, a transverse voltage can be generated by a longitudinal temperature gradient. This thermoelectric counterpart of the Anomalous Hall effect (AHE) is dubbed the  Anomalous Nernst effect (ANE). Expected to scale with spontaneous magnetization, both these effects arise because of the Berry curvature at the Fermi energy. Here, we report the observation of a giant ANE in a newly-discovered magnetic Weyl semimetal Co$_3$Sn$_2$S$_2$ crystal. Hall resistivity and Nernst signal both show sharp jumps at a threshold field and exhibit a clear hysteresis loop below the ferromagnetic transition temperature. The ANE signal peaks a maximum value of $\sim$ 5 $\mu$V/K which is comparable to the largest seen in any magnetic material. Moreover, the anomalous transverse thermoelectric conductivity $\alpha_{yx}$ becomes as large as $\sim$ 10 A/K.m at 70 K, the largest in known semimetals. The observed ANE signal is much larger than what is expected according to the magnetization.
%suggesting a new mechanism in Co$_3$Sn$_2$S$_2$ distinct to
%conventional ferromagnetic metals
%We found that the anomalous Hall conductivity is near a constant ($\backsim$ 1300 ($\Omega cm)^{-1}$) against temperature ($< $100K) and charge conductivity, implying a intrinsic Berry curvature mechanism. More importantly,

\end{abstract}

\maketitle

\section{\label{sec:level1}Introduction}

The Nernst effect, the transverse electric field generated by a
longitudinal thermal gradient in the presence of a magnetic field,
has triggered renewed attention in condense matter physics since the
discovery of the pseudogap phase in
cuprates\cite{ZhuAnXuNernst,PRBYaYWang}. Conductors with a large
Nernst coefficient are important for device applications as in
cryogenic refrigerations\cite{freibert2001thermomagnetic}, but such
devices have not been realized in practice because of their low
conversion efficiency. Recent years, the large Nernst effect has
been observed in correlated electron systems\cite{behnia2009nernst,CeCoIn,PRBURuSi},
conventional semimetals\cite{behnia2007nernst}, as well as in
metallic ferromagnets\cite{PRLPureFe,PRLCuCrSe,Fe3O4}. Consequently, a number
of novel ground states\cite{behnia2009nernst} and exotic electronic
orders\cite{PRlPrFeP,PRBURuSi,Stripephase} can be identified by
measuring the Nernst effect. For some ferromagnetic metals, in
particular, the Nernst signal was observed below $T_c$ even in the
absence of external magnetic field\cite{PRLPureFe,PRLCuCrSe,Fe3O4}.
This phenomenon, known as the anomalous Nernst effect (ANE),  is
observed to be proportional to the magnetization.
The underlying physics is that the spontaneous magnetization in
these materials plays a role of the intrinsic magnetic field,
geometrically connected to the Berry curvature of the Bloch bands at
the Fermi energy\cite{xiao2006berry}.

In recent years topological Dirac/Weyl semimetal materials have been
theoretically predicted and experimentally discovered. The
electronic structures of these materials have the topologically
robust\cite{RevModPhys,PRBWan,PRlyong,PRBYang,PRLDai} and
symmetry-protected bulk energy bands which linearly intersect at
some special points (the Dirac points) or symmetry axis near the
Fermi level\cite{PRBnode,PhysRevLettXu}. The breaking of inversion
symmetry or time-reversal symmetry (TRS) can split a Dirac point
into a pair of the Weyl points with opposite chiralities. The chiral
Weyl points are then the source or sink of the Berry curvature
$\Omega (k)$\cite{PRBQHS}, meaning that the Berry curvature is
singular at these points. So far, dozens of Dirac/Weyl semimetals
have been investigated
\cite{PhysRevBCdAs,liang2015ultrahigh,wang2012dirac,weng2015weyl,arnold2016negative,NbP,NbAs,zhang2017electron}
%and investigated with naturally broken
%inversion symmetry or field-induced broken TRS. associated
%with unconventional transport behavior associated with the chiral
%anomaly such as the negative longitudinal magnetoresistance (MR).
and several unique physical phenomena such as the large
magneto-resistivity (MR)\cite{zhang2017electron} and ultrahigh
mobility\cite{liang2015ultrahigh} have been observed. However, the
defining properties of such topological semimetals, including the Fermi
arc in surface states and chiral anomaly in charge transports, are
not easy to be identified experimentally.
%Therefore, it is very interesting to study
%the Nernst effect in this class of topological materials.
So far, a direct evidence of the Fermi arc has been clarified by
ARPES measurements in the
TaAs-family\cite{huang2015weyl,xu2015discovery}. While, a possible
indirect signature of the chiral anomaly is associated with the
negative longitudinal MR as investigated in several relevant
materials\cite{NegativeMR,arnold2016negative,li2016}. But some extrinsic factors
such as the current jetting and crystal
inhomogeneity\cite{CurrentJE} are hardly ruled out.
%resulting in an open issue for the negative MR in TDS/TWS.

Quite interestingly, there are two kinds of transverse transport
properties, the anomalous Hall effect (AHE) and the anomalous Nernst
effect (ANE), can help to probe the topological nature of charge
carriers in the ferromagnetic Weyl semimetals. This is because both
the transverse transport properties are contributed from the
intrinsic magnetic field in the occupied bands and thus deemed as
strong proofs of the finite Berry curvature originating from the
separation of Weyl nodes. Recently, a magnetic Weyl semimetal
Co$_3$Sn$_2$S$_2$ with a ferromagnetic kagome-lattice has been
reported to show an intrinsic anomalous Hall
effect\cite{MPI,wangRM}, but no experiment on the Nernst effect in
this material has been reported yet. It should be noticed that
unlike the Hall effect where the normal contribution in a metal is
always finite, the Nernst effect generally vanishes in ordinary
metals due to the, and thus the anomalous contribution may become very prominent.
On the other hand, a large AHE is not
necessary to cause a large ANE. This is because the AHE is determined by the
integration of the Berry curvature from all occupied bands,
while the ANE is governed by the Berry curvature at the Fermi
level\cite{xiao2010berry,xiao2006berry}. Thus, studying the ANE
is highly useful to confirm the contribution of the Berry curvature
and in turn verify the intrinsic Weyl state in a Weyl semimetal.

In this paper, we systematically study the ANE in the magnetic Weyl
semimetal Co$_3$Sn$_2$S$_2$. We find that the ANE signal reaches a maximum
value of $\sim$ 5 $\mu$ V/K at 70 K, yielding a giant transverse thermoelectrical conductivity of $\sim$ 10 A/K.m,
much larger than those of known ferromagnetic metals. This result shows that Co$_3$Sn$_2$S$_2$ is an
idea material candidate for future device application in cryogenic
refrigerations. Our study also provide insights in understanding the
intrinsic Weyl state and the correlation between AHE and ANE.

\section{\label{sec:level1}Results and Discussion}
%A semiclassical model suggests that the ratio of electronic mobility over Fermi energy contributes to the large Nernst effect, like semimetal Bi. However, the compensated metal\cite{PRLNbSe2},
\begin{figure}
\includegraphics[width=8cm]{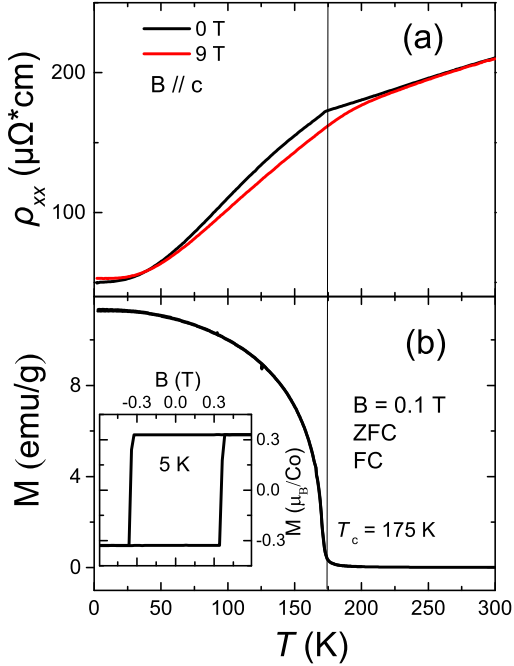}
\caption{ {a} Temperature dependence of the longitudinal electric resistivity($\rho_{xx}$) of Co$_3$Sn$_2$S$_2$ sample under zero field and 9 T. {b} Temperature dependence of magnetization with the FC modes at B $=$ 0.1 T. The inset shows the saturated magnetization vs. fields at 5 K.}
\end{figure}

Co$_3$Sn$_2$S$_2$ crystallizes in a rhombohedral structure with the
space group of R-3m. Co atoms form a kagome lattice with
corner-sharing triangles of Co atoms in the CoSn-layer, and the kagome
lattices in different layers are stacked along the c-axis with a corner-sharing
octahedra. The sample shows a magnetic moment of $\sim$
0.33$\mu_B$/Co along the c-axis with a Neel temperature of about 175
K in figure 1b. The FC and ZFC curves overlap each other very well
at 0.1 T, implying that the magnetic domains have been arranged
along c-axis by the small external fields. The clear hysteresis loop
with the saturated magnetization M$_s$ of 0.33 $\mu_B$/Co at 5 K is
observed in the inset of figure 1b, cosistent with the theoretical
prediction\cite{MPI,wangRM}. Temperature dependence of longitudinal
resistivity $\rho_{xx}$ and thermopower $S_{xx}$ of
Co$_3$Sn$_2$S$_2$ single crystal are shown in figure 1a. Overall the
resistivity shows a metallic behavior in the whole temperature
regime but exhibits an unexpected kink associated with a
ferromagnetic transition(FM) at about $T_c = $175 K, similar to the
previous report\cite{PhysRevB.96.014429}. As a magnetic field up to
9 T is applied, the resistivity does not change too much at high
temperature but starts to decrease below 200 K. The kink in
$\rho_{xx}(T)$ near $T_c$ broadens and becomes very smooth under 9
T, as seen in the literatures\cite{MPI,wangRM}.

%The zero-field thermopower is negative and shows a strong temperature-dependence in
%the measuring temperature range. With cooling temperature from room
%temperature, it exhibits a linear behavior and then shows the
%similar kink near $T_c$. A local minimum in $S_{xx}(T)$ at about 50
%K can be observed. The $S_{xx}$ is about 100 $\mu$ V/K at 300 K, two times larger than what was previously reported in the
%polycrystal\cite{Seebeck}. The applied magnetic field (9 T) does
%almost unchange the $S_{xx}$ except that it tremendously suppresses
%the kink in $S_{xx}(T)$, consistent with the behavior of resistivity.

%DEPENDENCE%%%%%%%%%%%%%%%%%%%%%%%%%%%%%%%%%%%%%%%%%%%%%%%%%%%%%%%%%%%%%%%%%%%%%%%%%%%%%%%%%%%%%%%%%%%%%%%%%%%%%%%%%%%%%%%%%%%%%%%%%%%%%%%%%%%%%%%%%%%%%%%%%%%%%%%%%%%%%%%%%%%%%%%%%

\begin{figure}
\includegraphics[width=8cm]{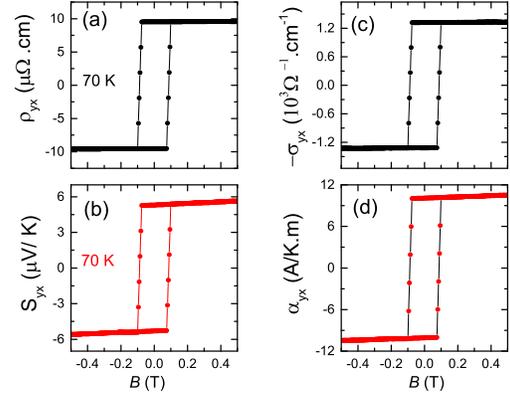}
\caption{Field dependence of transport coefficients at 70 K as B $\parallel$ c-axis $\perp$ I. (a) Hall resistivity $\rho_{yx}$ vs. magnetic fields. (b) The Nernst signal S$_{yx}$. (c) The Hall conductivity $\sigma_{yx}$, extracted from $\rho_{xx}$ and $\rho_{yx}$. (d) Transverse thermoelectric conductivity $\alpha_{yx}$, extracted from $\rho_{xx}$, $\rho_{yx}$, S$_{xx}$ and S$_{yx}$. The hysteretic behavior is clearly  observed at 70 K.}
\end{figure}

Figure 2a and 2c display the Hall resistivity and conductivity vs. the applied magnetic fields $B \parallel c \perp I$ at a representative temperature of 70 K. The $\rho_{yx}(B)$ significantly exhibits a steep rectangular hysteretic jump at low fields, and then becomes almost flat at a higher field. Its magnitude is comparable to what was reported previously\cite{MPI,wangRM} and is reversible with a field of 0.15 T at 70 K. The largest coercive field is about 0.7 T at 5 K (See figure S3 in SI). The Nernst signal $S_{yx}$ (figure 2b) shows a similar jump of about 10 $\mu$ V/K, which is over one order of magnitude larger than those of typical materials for the ANE such as Mn$_3$Sn\cite{nayak2016large}, pure metal Fe\cite{Mn3SnZHu} CuCr$_2$Se$_{4-x}$Br$_x$\cite{PRLCuCrSe} and single crystal Fe$_3$O$_4$\cite{Fe3O4}. The large S$_{yx}$ is also verified by the large Nernst angle $\theta_N = \mid S_{xy}/S_{xx}\mid$ of 16$\%$ (Figure 3b). Note that both Hall resistivity and Nernst signal show nearly the same magnetic field dependence as the magnetization curve (See figure S2 in SI), implying that the contribution of the anomalous Hall and Nernst effects dominates and the normal contribution (proportional to B) is negligibly small at low fields.

Field dependence of the Hall conductivity $-\sigma_{yx}$ and the transverse thermoelectric conductivity $\alpha_{yx}$ at 70 K is shown in figure 2c and 2d. The $\sigma_{yx}$ and $\alpha_{yx}$ can be written as the formulas $\sigma_{yx} = - \rho_{yx}/(\rho_{yx}^2+\rho^2)$ and $\alpha_{yx} = (\rho_{xx}S_{yx}-\rho_{yx}S_{xx})/\rho_{xx}^2$\cite{Mn3SnZHu}. The obtained $\mid\sigma_{yx}\mid$ at the saturated field 0.5 T reaches $\sim$ 1320 ($\Omega cm)^{-1}$, which is very close to the calculated result (1310 ($\Omega cm)^{-1}$) from the integral of Berry curvature along $k_{z}$ in the system \cite{MPI,wangRM}. Similar to the large $\sigma_{yx}$, the $\alpha_{yx}$ is found to be very large, reaching approximately 11 A/K.m at 70 K. This value is one or two orders magnitude of larger than the other typical ferromagnets or Mn$_3$Sn with $\alpha_{yx} \sim$ 0.01 - 1 A/K.m, as shown in Figure 4b.

Temperature dependence of the AHR, $\rho_{yx}^{A}$, (extrapolating the high-field part of $\rho_{yx}$ back to zero field) and AHC, $\sigma_{xy}^A$, are illustrated in figure 3a. The $\rho_{yx}^{A}$ shows a strong temperature-dependence below $T_c$, and peaks a maximum value of 21 $\mu \Omega cm$ around 140 K. While the $\sigma_{xy}^A$ is almost temperature-independence below 100 K. Such feature is also observed in pure materials such as Fe, Co, Ni and Gd, where the mechanism of the AHE has been suggested to be the intrinsic Berry-phase mechanism\cite{PRLPureFe}. We plot the $\sigma_{xy}^A$ as a function of $\sigma_{xx}$ in the main plane of figure 3b. It describes that $\sigma_{xy}^A$ is weakly dependent of $\sigma_{xx}$ at low temperature regime( $<$ 100 K), similar to the case in Fe, Ni pure metals\cite{PRLPureFe} (See SI). A large anomalous Hall angle $\mid\sigma_{xy}^A/\sigma\mid$ reaches a maximum value about 15$\%$ at 150 K and gradually decreases to 6.3$\%$ at 5 K, as shown in the inset of figure 3b. The large Hall angle in Co$_3$Sn$_2$S$_2$ is close to the value of 16$\%$ in GdPtBi, a typical field-induced Weyl-semimetal\cite{GdBiPt}, but is much larger than that of the noncolinear antiferromagnetic Mn$_3$Sn (3.2$\%$)\cite{Mn3SnNernst} and Mn$_3$Ge (5$\%$ )\cite{nayak2016large}.

%The saturated fields $B_H^0$ as a function of temperature is shown in the
\begin{figure}
\includegraphics[angle=0,width=8cm,clip]{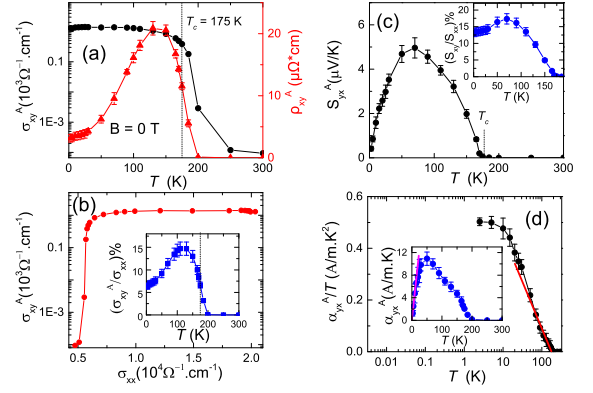}
\caption{(a) Temperature dependence of the anomalous Hall conductivity $\sigma_{xy}^A$ and Hall resistivity $\rho_{yx}^{A}$ at zero magnetic field. (b) $\sigma_(xy)^A$ as a function of $\sigma_(xx)$ in the main plane. The inset shows temperature dependence of the Hall angle ($\mid\sigma_(xy)^A/\sigma\mid$) at zero magnetic field.(c) Temperature dependence of the anomalous Nernst signal S$_{yx}^A$. The inset shows the Nernst angle $\mid\theta_N = S_{xy}/S_{xx}\mid$ as a function of temperature. (d) Temperature dependence of the $\alpha_{yx}/T$. The inset shows the transverse thermoelectric conductivity $\alpha_{yx}$ vs. temperature. }
\end{figure}

The ANE signal $S_{yx}^A$ as a function of temperature is displayed in figure 3c. The $S_{yx}^A$ strongly depends on temperature, rapidly increases below $T_c$ and then peaks at around 70 K, a half of the peak temperature of $\rho_{yx}^{A}$ in figure 3a. Similar to the large Hall angle, the large $S_{yx}^A$ can be verified by the Nernst angle $\theta_N = \mid S_{xy}/S_{xx}\mid$ in the inset of figure 3c. The value reaches 17\% at 70 K, which is comparable with the previous findings in the Co$_2$MnGa system\cite{CoMnGa}, a FM semimetal. Correspondingly, temperature dependence of the $\alpha_{yx}^A$ is mapped in the inset of figure 3d, and the $\alpha_{yx}^A$ rapidly increases below 200 K, showing a maximum value of $\sim$ 11 A/K.m at 50 K, followed by a monotonous decrease to low temperatures. Such large $\alpha_{yx}^A$ is one or two orders magnitude larger than the other ferromagnets with a small $\alpha_{yx}$ $\sim$ 0.01-1 A/K.m (See figure S4 in SI). Note that the $\alpha_{yx}^A$ is roughly obey the linear T-dependence at low temperatures, but closely follow the relation $\propto T\log(T)$ at a temperature regime from 70 K to 175 K, as shown in figure 3d. Similar behavior has been reported in Co$_2$MnGa system\cite{CoMnGa}.

%but at the high fields or the high temperatures the normal Nernst signal almost completely dominates because of a good linear component.

%FIELD DEPENDANCE%%%%%%%%%%%%%%%%%%%%%%%%%%%%%%%%%%%%%%%%%%%%%%%%%%%%%%%%%%%%%%%%%%%%%%%%%%%%%%%%%%%%%%%%%%%%%%%%%%%%%%%%%%%%%%%%%%%%%%%%%%%%%%%%%%%%%%%%%%%%%%%%%%%%%%%%%%%%%%%%%%%%%%%%%%%%%%
\begin{figure}
\includegraphics[angle=0,width=8cm,clip]{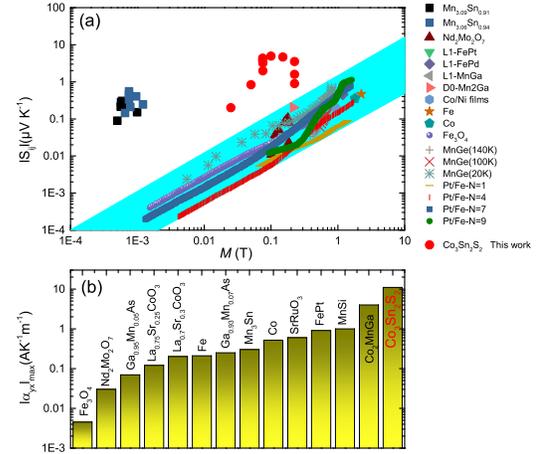}
\caption{(a) Magnetization dependence of the spontaneous Nernst effect for FM metals, Mn$_3$Sn and Co$_3$Sn$_2$S$_2$. (b) The transverse thermoelectric conductivity $|\alpha_{yx}|$ vs. the various ferromagnets, Mn$_3$Sn and Co$_3$Sn$_2$S$_2$. All those ANE results are obtained for various ferromagnets below their Curie temperatures(Methods).}
\end{figure}

The significant difference between the giant ANE in Co$_3$Sn$_2$S$_2$, Mn$_3$Sn and in ferromagnets is demonstrated in figure 4a. We plot the ANE as functions of the magnetization with a logarithmic scale for various ferromagnetic metals, Mn$_3$Sn and Co$_3$Sn$_2$S$_2$ (See Methods). The ANE in ferromagnets is known bo be roughly proportional to the magnetization $M$, $|S_{yx}| = |Q_s| \mu_0 M$, where $|Q_s|$ is the anomalous Nernst coefficient. It can be seen that almost all of ferromagnetic metals follow this relation and their ANE signal becomes more larger with increasing magnetization. The shaded region covers all of the data points for ferromagnets. An exception for the antiferromagnetic Mn$_3$Sn observed strongly violates this scaling, which is ascribed to the enhanced Berry curvature at the Fermi energy\cite{Mn3SnNernst}. Similarly, the Co$_3$Sn$_2$S$_2$ does not follow this relation. The $S_{yx}$ is concretely far away from this trend and more than one and two order magnitude larger than what would be expected based on the scaling relation. In addition, the intrinsic transverse thermoelectric conductivity $|\alpha_{yx}|$ is close to 11 A/K.m for Co$_3$Sn$_2$S$_2$, the largest value compared to the other ferromagnets and semimetals as shown in figure 4b. The observed large $|\alpha_{yx}|$ in the Weyl magnets are potentially useful to realize the thermoelectric devices for cryogenic refrigeration.

The mechanism of the significantly large intrinsic ANE in Co$_3$Sn$_2$S$_2$ should be distinct from the conventional one for ferromagnets. In the formwork of the Boltzmann approach, the anomalous Hall conductivity $\sigma_{yx}^A$ and the anomalous transverse thermoelectric conductivity $\alpha_{yx}^A$ can be written as the formulas\cite{xiao2010berry}: $\sigma_{yx}^A = -\frac{e^2}{\hbar}\int\frac{dk}{2\pi^3}\Omega_{n,z}(k)f_{nk}$, and
$\alpha_{yx}^A = -\frac{e}{T\hbar}\int\frac{dk}{2\pi^3}\Omega_{n,z}(k){(\varepsilon_{n,k}-\mu)f_{n,k}+k_B T ln[1+e^{-\beta(\varepsilon_{n,k}-\mu)}]}$
Here, $\Omega_{n,k}$ is the Berry curvature along the $z$ direction, and $f_{nk} = f(E_{nk})$ is the Fermi distribution function with the band index $n$ and the wave vector $k$. $\varepsilon_{n,k}$ and $\beta$ donate the band energy and $1/k_B T$. From the equations, the $\alpha_{yx}^A$ is governed by the Berry curvature around the Fermi energy, while the $\sigma_{yx}^A$ is given by the summation of the Berry curvature for all the occupied bands. According to these equations, the $\alpha_{yx}^A$ at low temperatures can be reduced to the Mott relations $\alpha_{yx}^A = \frac{\pi^2 k_B^2 T}{4e}$$\frac{\partial \sigma_{yx}^A}{\partial \varepsilon}$. Our result in $\alpha_{yx}^A(T)$ showing a roughly linear temperature-dependence at low temperatures is consistent with the Mott rule.

%Indeed, a recent calculation found that Mn3Sn may have a large anomalous Hall conductivity20, which can be estimated
%by integrating the Berry curvature of the occupied bands over the entire Brillouin zone9.More recently, the possibility of aWeyl metal
%has been proposed, where the bands crossing EF have several Weyl points, around which the Berry curvature diverges

%By integration by parts, the $\alpha_{yx}^A$ and $\sigma_{yx}^A$ can be related as follows:

%\alpha_{yx}^A = -\frac{1}{e}\int\frac{\partial f}{\partial \mu}\sigma_{yx}^A(\varepsilon)\frac{\epsilon-\mu}{\emph{T}}{d\varepsilon}

\section{\label{sec:level1}Conclusion}
%CONCLUSION%%%%%%%%%%%%%%%%%%%%%%%%%%%%%%%%%%%%%%%%%%%%%%%%%%%%%%%%%%%%%%%%%%%%%%%%%%%%%%%%%%%%%%%%%%%%%%%%%%%%%%%%%%%%%%%%%%%%%%%%%%%%%%%%%%%%%%%%%%%%%%%%%%%%%%%%%%%%%%%%%%%%%%%%%%%%%%%%%%%%%%%%%
In summary, we report an observation of the giant ANE and the large intrinsic transverse thermoelectric conductivity in the magnetic Weyl semimetal Co$_3$Sn$_2$S$_2$. The Nernst signal $S_{yx}$ reaches $\sim$ 5 $\mu$ V/K at 70 K, much larger than those of known ferromagnetic metals. Furthermore, the $\alpha_{xy}^A \sim 10 A/K.m$ in the sample is over one or two order magnitude larger than those of other ferromagnets where $\alpha_{xy}^A = 0.01 -1$ A/K/m. Our experimental results indicate that the giant ANE in Co$_3$Sn$_2$S$_2$ can be ascribed to the enhanced contribution from the Berry curvature very close to Fermi levels. The enhanced $\alpha_{xy}$ in Weyl magnets paves a path to realize potentially thermopile devices for the thermoelectric power generation. Our experimental results also highlight the complementary roles of ANE and AHE in revealing the intrinsic Weyl state in the magnetic Weyl semimetals.

\section{\label{sec:level1}Method}

Large single crystals of Co$_3$Sn$_2$S$_2$ with the minimeter size were grown
through the Bridgman technique, as reported in the previous reports\cite{PhysRevB.96.014429}.

The single crystal X-ray diffraction were obtained using a D/Max-rA diffractometer with CuK$_{\alpha}$
radiation and a graphite monochromator at the room temperature, which determines the crystal grown
orientation. The composition of the crystals were obtained by energy
dispersive X-ray (EDX) spectroscopy. The sample is polished and then cut into a  bar-shape with a size of 3.5$\times$2$\times$0.2 mm$^3$.
The magnetization measurements were done using a commercial SQUID magnetometer.
The (magneto)resistivity and Hall coefficient measurements were performed using the standard
four-terminal method in a commercial Quantum Design PPMS-9 system.
The thermal power measurement were performed with a one-heater-two-thermometers technique in PPMS with
a high-vacuum environment. Two Chromel-constantan (type E) thermocouples were employed to measure the
temperature difference generated by a small heater chip.

In figure4a, the Nernst signal $S_{yx}$ for various ferromagnets below the Curie temperatures is abstracted the ANE results, as reported in those
literatures. For example Fe(300 K)\cite{perdew1996}; Co(300K)\cite{perdew1996}; Fe$_3$O$_4$(300K)\cite{Fe3O4}; MnGe(140K)\cite{vanderbilt1990}, MnGe(100K)\cite{vanderbilt1990} and MnGe(20K)\cite{vanderbilt1990}; Nd$_2$Mo$_2$O$_7$\cite{hanasaki2008};
Co/Ni film(300K)\cite{hasegawa2015}; Pt/Fe multilayer(300K)\cite{mostofi2008}; MnGa(300K)\cite{hasegawa2015}; Mn$_2$Ga(300K)\cite{hasegawa2015}; FePt(300K)\cite{hasegawa2015}; and FePd(300K)\cite{hasegawa2015}; Mn$_3$Sn\cite{Mn3SnNernst}

In figure 4b, the maximum $|\alpha_{xy}^A|$ values are taken from each magnetic material, Mn$_3$Sn and Co$_2$MnGa.
$|\alpha_{xy}^A| \sim 1 A/Km$ for MnSi\cite{hirokane2016}, $|\alpha_{xy}^A| \sim 0.6 A/Km$ for SrRuO$_3$\cite{PRLPureFe}, $|\alpha_{xy}^A| \sim 0.3 A/Km$ for Mn$_3$Sn\cite{Mn3SnNernst}, $|\alpha_{xy}^A| \sim$ 0.2 and 0.12 A/Km for La$_{1-x}$Sr$_x$CoO$_3$(x=0.3 and 0.25)\cite{PRLPureFe}, $|\alpha_{xy}^A| \sim 0.03 A/Km$ for Nd$_2$Mo$_2$O$_7$\cite{hanasaki2008}, $|\alpha_{xy}^A| \sim 0.25 and 0.07 A/Km$ for Ga$_{1-x}$Mn$_x$Si\cite{pu2008mott}, $|\alpha_{xy}^A| \sim 0.0045 A/Km$ for Fe$_3$O$_4$\cite{Fe3O4}, and $|\alpha_{xy}^A| \sim$ 0.9, 0.52 and 0.21 A/Km for FePt, Co and Fe\cite{weischenberg2013}, $|\alpha_{xy}^A| \sim 4 A/Km$ for Co$_2$MnGa\cite{CoMnGa}, and $|\alpha_{xy}^A| \sim 11 A/Km$ for Co$_3$Sn$_2$S$_2$.

\bibliographystyle{unsrt}
\bibliography{CoSnSNC}

\begin{thebibliography}{10}

\bibitem{ZhuAnXuNernst}
ZA~Xu, NP~Ong, Yayu Wang, T~Kakeshita, and S~Uchida.
\newblock Vortex-like excitations and the onset of superconducting phase
  fluctuation in underdoped ${La}_{2-x}{Sr}_x {CuO}_ 4$.
\newblock {\em Nature}, 406(6795):486, 2000.

\bibitem{PRBYaYWang}
Yayu Wang, Lu~Li, and N.~P. Ong.
\newblock Nernst effect in high-${T}_{c}$ superconductors.
\newblock {\em Phys. Rev. B}, 73:024510, Jan 2006.

\bibitem{freibert2001thermomagnetic}
Franz Freibert, Timothy~W Darling, Albert Migliori, and Stuart~A Trugman.
\newblock Thermomagnetic effects and measurements.
\newblock In {\em Semiconductors and Semimetals}, volume~70, pages 207--244.
  Elsevier, 2001.

\bibitem{behnia2009nernst}
Kamran Behnia.
\newblock The nernst effect and the boundaries of the fermi liquid picture.
\newblock {\em Journal of Physics: Condensed Matter}, 21(11):113101, 2009.

\bibitem{CeCoIn}
R~Bel, K~Behnia, Y~Nakajima, K~Izawa, Y~Matsuda, H~Shishido, R~Settai, and
  Y~Onuki.
\newblock Giant nernst effect in {CeCoIn$_5$}.
\newblock {\em Physical Review Letters}, 92(21):217002, 2004.

\bibitem{PRBURuSi}
Romain Bel, Hao Jin, Kamran Behnia, Jacques Flouquet, and Pascal Lejay.
\newblock Thermoelectricity of ${U}{Ru}_{2}{Si}_{2}$: Giant nernst effect in
  the hidden-order state.
\newblock {\em Phys. Rev. B}, 70:220501, Dec 2004.

\bibitem{behnia2007nernst}
Kamran Behnia, Marie-Aude M{\'e}asson, and Yakov Kopelevich.
\newblock Nernst effect in semimetals: The effective mass and the figure of
  merit.
\newblock {\em Physical Review Letters}, 98(7):076603, 2007.

\bibitem{PRLPureFe}
T.~Miyasato, N.~Abe, T.~Fujii, A.~Asamitsu, S.~Onoda, Y.~Onose, N.~Nagaosa, and
  Y.~Tokura.
\newblock Crossover behavior of the anomalous hall effect and anomalous nernst
  effect in itinerant ferromagnets.
\newblock {\em Phys. Rev. Lett.}, 99:086602, Aug 2007.

\bibitem{PRLCuCrSe}
Wei-Li Lee, S.~Watauchi, V.~L. Miller, R.~J. Cava, and N.~P. Ong.
\newblock Anomalous hall heat current and nernst effect in the
  ${CuCr}_{2}{Se}_{4-x}{Br}_{x}$ ferromagnet.
\newblock {\em Phys. Rev. Lett.}, 93:226601, Nov 2004.

\bibitem{Fe3O4}
R.~Ramos, M.~H. Aguirre, A.~Anad\'on, J.~Blasco, I.~Lucas, K.~Uchida, P.~A.
  Algarabel, L.~Morell\'on, E.~Saitoh, and M.~R. Ibarra.
\newblock Anomalous nernst effect of ${Fe}_{3}{O}_{4}$ single crystal.
\newblock {\em Phys. Rev. B}, 90:054422, Aug 2014.

\bibitem{PRlPrFeP}
A.~Pourret, K.~Behnia, D.~Kikuchi, Y.~Aoki, H.~Sugawara, and H.~Sato.
\newblock Drastic change in transport of entropy with quadrupolar ordering in
  ${PrFe}_{4}{P}_{12}$.
\newblock {\em Phys. Rev. Lett.}, 96:176402, May 2006.

\bibitem{Stripephase}
Francis Lalibert¨¦ David LeBoeuf Nicolas Doiron-Leyraud J. Chang J. Q. Yan
  J.-G. Cheng J.-S. Zhou J. B. Goodenough S. Pyon T. Takayama H. Takagi Y.
  Tanaka Louis~Taillefer Olivier Cyr-Choini¨¨re, R.~Daou.
\newblock Enhancement of the nernst effect by stripe order in a high-tc
  superconductor.
\newblock {\em Nature}, 458:743, 2009.

\bibitem{xiao2006berry}
Di~Xiao, Yugui Yao, Zhong Fang, and Qian Niu.
\newblock Berry-phase effect in anomalous thermoelectric transport.
\newblock {\em Physical Review Letters}, 97(2):026603, 2006.

\bibitem{RevModPhys}
N.~P. Armitage, E.~J. Mele, and Ashvin Vishwanath.
\newblock Weyl and dirac semimetals in three-dimensional solids.
\newblock {\em Rev. Mod. Phys.}, 90:015001, Jan 2018.

\bibitem{PRBWan}
Xiangang Wan, Ari~M. Turner, Ashvin Vishwanath, and Sergey~Y. Savrasov.
\newblock Topological semimetal and fermi-arc surface states in the electronic
  structure of pyrochlore iridates.
\newblock {\em Phys. Rev. B}, 83:205101, May 2011.

\bibitem{PRlyong}
S.~M. Young, S.~Zaheer, J.~C.~Y. Teo, C.~L. Kane, E.~J. Mele, and A.~M. Rappe.
\newblock Dirac semimetal in three dimensions.
\newblock {\em Phys. Rev. Lett.}, 108:140405, Apr 2012.

\bibitem{PRBYang}
Kai-Yu Yang, Yuan-Ming Lu, and Ying Ran.
\newblock Quantum hall effects in a weyl semimetal: Possible application in
  pyrochlore iridates.
\newblock {\em Phys. Rev. B}, 84:075129, Aug 2011.

\bibitem{PRLDai}
Chen Fang, Matthew~J. Gilbert, Xi~Dai, and B.~Andrei Bernevig.
\newblock Multi-weyl topological semimetals stabilized by point group symmetry.
\newblock {\em Phys. Rev. Lett.}, 108:266802, Jun 2012.

\bibitem{PRBnode}
A.~A. Burkov, M.~D. Hook, and Leon Balents.
\newblock Topological nodal semimetals.
\newblock {\em Phys. Rev. B}, 84:235126, Dec 2011.

\bibitem{PhysRevLettXu}
Gang Xu, Hongming Weng, Zhijun Wang, Xi~Dai, and Zhong Fang.
\newblock Chern semimetal and the quantized anomalous hall effect in
  ${HgCr}_{2}{Se}_{4}$.
\newblock {\em Phys. Rev. Lett.}, 107:186806, Oct 2011.

\bibitem{PRBQHS}
Shuichi Murakami and Shun-ichi Kuga.
\newblock Universal phase diagrams for the quantum spin hall systems.
\newblock {\em Phys. Rev. B}, 78:165313, Oct 2008.

\bibitem{PhysRevBCdAs}
Zhijun Wang, Hongming Weng, Quansheng Wu, Xi~Dai, and Zhong Fang.
\newblock Three-dimensional dirac semimetal and quantum transport in
  cd${}_{3}$as${}_{2}$.
\newblock {\em Phys. Rev. B}, 88:125427, Sep 2013.

\bibitem{liang2015ultrahigh}
Tian Liang, Quinn Gibson, Mazhar~N Ali, Minhao Liu, RJ~Cava, and NP~Ong.
\newblock Ultrahigh mobility and giant magnetoresistance in the dirac semimetal
  {Cd}$_3${As}$_2$.
\newblock {\em Nature materials}, 14(3):280--284, 2015.

\bibitem{wang2012dirac}
Zhijun Wang, Yan Sun, Xing-Qiu Chen, Cesare Franchini, Gang Xu, Hongming Weng,
  Xi~Dai, and Zhong Fang.
\newblock Dirac semimetal and topological phase transitions in {A}$_3${Bi} ({A
  = Na, K, Rb}).
\newblock {\em Physical Review B}, 85(19):195320, 2012.

\bibitem{weng2015weyl}
Hongming Weng, Chen Fang, Zhong Fang, B~Andrei Bernevig, and Xi~Dai.
\newblock Weyl semimetal phase in noncentrosymmetric transition-metal
  monophosphides.
\newblock {\em Physical Review X}, 5(1):011029, 2015.

\bibitem{arnold2016negative}
Frank Arnold, Chandra Shekhar, Shu-Chun Wu, Yan Sun, Ricardo~Donizeth Dos~Reis,
  Nitesh Kumar, Marcel Naumann, Mukkattu~O Ajeesh, Marcus Schmidt, Adolfo~G
  Grushin, et~al.
\newblock Negative magnetoresistance without well-defined chirality in the weyl
  semimetal {TaP}.
\newblock {\em Nature communications}, 7, 2016.

\bibitem{NbP}
Chandra Shekhar, Ajaya~K Nayak, Yan Sun, Marcus Schmidt, Michael Nicklas, Inge
  Leermakers, Uli Zeitler, Yurii Skourski, Jochen Wosnitza, Zhongkai Liu,
  et~al.
\newblock Extremely large magnetoresistance and ultrahigh mobility in the
  topological weyl semimetal candidate {NbP}.
\newblock {\em Nature Physics}, 11(8):645--649, 2015.

\bibitem{NbAs}
Nirmal~Jeevi Ghimire, Yongkang Luo, Madhab Neupane, DJ~Williams, ED~Bauer, and
  F~Ronning.
\newblock Magnetotransport of single crystalline {NbAs}.
\newblock {\em Journal of Physics: Condensed Matter}, 27(15):152201, 2015.

\bibitem{zhang2017electron}
Cheng-Long Zhang, Zhujun Yuan, Qing-Dong Jiang, Bingbing Tong, Chi Zhang,
  XC~Xie, and Shuang Jia.
\newblock Electron scattering in tantalum monoarsenide.
\newblock {\em Physical Review B}, 95(8):085202, 2017.

\bibitem{huang2015weyl}
Shin-Ming Huang, Su-Yang Xu, Ilya Belopolski, Chi-Cheng Lee, Guoqing Chang,
  BaoKai Wang, Nasser Alidoust, Guang Bian, Madhab Neupane, Chenglong Zhang,
  et~al.
\newblock A weyl fermion semimetal with surface fermi arcs in the transition
  metal monopnictide {TaAs} class.
\newblock {\em Nature communications}, 6:7373, 2015.

\bibitem{xu2015discovery}
Su-Yang Xu, Ilya Belopolski, Nasser Alidoust, Madhab Neupane, Guang Bian,
  Chenglong Zhang, Raman Sankar, Guoqing Chang, Zhujun Yuan, Chi-Cheng Lee,
  et~al.
\newblock Discovery of a weyl fermion semimetal and topological fermi arcs.
\newblock {\em Science}, 349(6248):613--617, 2015.

\bibitem{NegativeMR}
Jun Xiong, Satya~K Kushwaha, Tian Liang, Jason~W Krizan, Max Hirschberger, Wudi
  Wang, RJ~Cava, and NP~Ong.
\newblock Evidence for the chiral anomaly in the dirac semimetal {Na}$_3${Bi}.
\newblock {\em Science}, 350(6259):413--416, 2015.

\bibitem{li2016}
Yuke Li, Lin Li, Jialu Wang, Tingting Wang, Xiaofeng Xu, Chuanying Xi, Chao
  Cao, and Jianhui Dai.
\newblock Resistivity plateau and negative magnetoresistance in the topological
  semimetal {TaSb}$_{2}$.
\newblock {\em Physical Review B}, 94(12):121115, 2016.

\bibitem{CurrentJE}
RD~Dos~Reis, MO~Ajeesh, N~Kumar, F~Arnold, C~Shekhar, M~Naumann, M~Schmidt,
  M~Nicklas, and E~Hassinger.
\newblock On the search for the chiral anomaly in weyl semimetals: The negative
  longitudinal magnetoresistance.
\newblock {\em New Journal of Physics}, 18(8):085006, 2016.

\bibitem{MPI}
Enke Liu, Yan Sun, Nitesh Kumar, Lukas Muechler, Aili Sun, Lin Jiao, Shuo-Ying
  Yang, Defa Liu, Aiji Liang, Qiunan Xu, et~al.
\newblock Giant anomalous hall effect in a ferromagnetic kagome-lattice
  semimetal.
\newblock {\em Nature Physics}, page~1, 2018.

\bibitem{wangRM}
Qi~Wang, Yuanfeng Xu, Rui Lou, Zhonghao Liu, Man Li, Yaobo Huang, Dawei Shen,
  Hongming Weng, Shancai Wang, and Hechang Lei.
\newblock Large intrinsic anomalous hall effect in half-metallic ferromagnet
  ${Co_3Sn_2S_2}$ with magnetic weyl fermions.
\newblock {\em Nature Communications}, 9(1):3681, 2018.

\bibitem{xiao2010berry}
Di~Xiao, Ming-Che Chang, and Qian Niu.
\newblock Berry phase effects on electronic properties.
\newblock {\em Reviews of Modern Physics}, 82(3):1959, 2010.

\bibitem{PhysRevB.96.014429}
Mohamed~A. Kassem, Yoshikazu Tabata, Takeshi Waki, and Hiroyuki Nakamura.
\newblock Low-field anomalous magnetic phase in the kagome-lattice shandite
  {Co$_3$Sn$_2$S$_2$}.
\newblock {\em Phys. Rev. B}, 96:014429, Jul 2017.

\bibitem{nayak2016large}
Ajaya~K Nayak, Julia~Erika Fischer, Yan Sun, Binghai Yan, Julie Karel,
  Alexander~C Komarek, Chandra Shekhar, Nitesh Kumar, Walter Schnelle,
  J{\"u}rgen K{\"u}bler, et~al.
\newblock Large anomalous hall effect driven by a nonvanishing berry curvature
  in the noncolinear antiferromagnet {Mn$_3$Ge}.
\newblock {\em Science advances}, 2(4):e1501870, 2016.

\bibitem{Mn3SnZHu}
Xiaokang Li, Liangcai Xu, Linchao Ding, Jinhua Wang, Mingsong Shen, Xiufang Lu,
  Zengwei Zhu, and Kamran Behnia.
\newblock Anomalous nernst and righi-leduc effects in ${Mn}_{3}{Sn}$: Berry
  curvature and entropy flow.
\newblock {\em Phys. Rev. Lett.}, 119:056601, Aug 2017.

\bibitem{GdBiPt}
T.~Suzuki, R.~Chisnell, A.~Devarakonda, Y.~T. Liu, W.~Feng, and D.~Xiao.
\newblock Large anomalous hall effect in a half-heusler antiferromagnet.
\newblock {\em Nature Physics}, 12:1119, 2016.

\bibitem{Mn3SnNernst}
Tomita Takahiro Koretsune Takashi Suzuki Michi-To Nishio-Hamane-Daisuke Arita
  Ryotaro Otani Yoshichika Nakatsuji~Satoru Ikhlas, Muhammad.
\newblock Large anomalous nernst effect at room temperature in a chiral
  antiferromagnet.
\newblock {\em Nature Physics}, 13:1085, 2017.

\bibitem{CoMnGa}
Akito Sakai, Yo~Pierre Mizuta, Agustinus~Agung Nugroho, Rombang Sihombing,
  Takashi Koretsune, Michi-To Suzuki, Nayuta Takemori, Rieko Ishii, Daisuke
  Nishio-Hamane, Ryotaro Arita, et~al.
\newblock Giant anomalous nernst effect and quantum-critical scaling in a
  ferromagnetic semimetal.
\newblock {\em Nature Physics}, page~1, 2018.

\bibitem{perdew1996}
John~P Perdew, Kieron Burke, and Matthias Ernzerhof.
\newblock Generalized gradient approximation made simple.
\newblock {\em Physical review letters}, 77(18):3865, 1996.

\bibitem{vanderbilt1990}
David Vanderbilt.
\newblock Soft self-consistent pseudopotentials in a generalized eigenvalue
  formalism.
\newblock {\em Physical Review B}, 41(11):7892, 1990.

\bibitem{hanasaki2008}
N~Hanasaki, K~Sano, Y~Onose, T~Ohtsuka, S~Iguchi, I~K{\'e}zsm{\'a}rki,
  S~Miyasaka, S~Onoda, N~Nagaosa, and Y~Tokura.
\newblock Anomalous nernst effects in pyrochlore molybdates with spin
  chirality.
\newblock {\em Physical review letters}, 100(10):106601, 2008.

\bibitem{hasegawa2015}
K~Hasegawa, M~Mizuguchi, Y~Sakuraba, T~Kamada, T~Kojima, T~Kubota, S~Mizukami,
  T~Miyazaki, and K~Takanashi.
\newblock Material dependence of anomalous nernst effect in perpendicularly
  magnetized ordered-alloy thin films.
\newblock {\em Applied Physics Letters}, 106(25):252405, 2015.

\bibitem{mostofi2008}
Arash~A Mostofi, Jonathan~R Yates, Young-Su Lee, Ivo Souza, David Vanderbilt,
  and Nicola Marzari.
\newblock wannier90: A tool for obtaining maximally-localised wannier
  functions.
\newblock {\em Computer physics communications}, 178(9):685--699, 2008.

\bibitem{hirokane2016}
Yuji Hirokane, Yasuhide Tomioka, Yoshinori Imai, Atsutaka Maeda, and Yoshinori
  Onose.
\newblock Longitudinal and transverse thermoelectric transport in mnsi.
\newblock {\em Physical Review B}, 93(1):014436, 2016.

\bibitem{pu2008mott}
Yong Pu, Daichi Chiba, Fumihiro Matsukura, Hideo Ohno, and Jing Shi.
\newblock Mott relation for anomalous hall and nernst effects in
  {Ga$_{1-x}$Mn$_x$As} ferromagnetic semiconductors.
\newblock {\em Physical review letters}, 101(11):117208, 2008.

\bibitem{weischenberg2013}
J{\"u}rgen Weischenberg, Frank Freimuth, Stefan Bl{\"u}gel, and Yuriy
  Mokrousov.
\newblock Scattering-independent anomalous nernst effect in ferromagnets.
\newblock {\em Physical Review B}, 87(6):060406, 2013.

\end{thebibliography}

\section*{Acknowledgements}
We thank K. Behnia for inspiring discussions and suggestions.
We also thank H. Wang, and J. Yang for some technical support.
This research was supported in part by the NSF of China (under
Grants No. 11474082) and the National Key Projects for Research and Development of China (Contract No. 2016YFA0300402).
Yu-Ke Li was supported by an open
program from Wuhan National High Magnetic Field Center (2016KF03).

\section*{Author contributions}

Y. Li designed the research. H. Yang synthesized the samples and performed the electronic and thermal transport measurements.
W. You performed the XRD measurements and analyzed the structure parameters. C. Cao performed the first-principle calculation.
C. Xi, and J. Wang assisted the measurements. C. Cao, M. L. Tian, Z. A. Xu, J. Dai, and Y. Li, discussed the data, interpreted the results. J. Dai and Y. Li wrote the paper.

\section*{Additional information}
Competing financial interests: The authors declare no competing financial interests.
Correspondence and requests for materials should be addressed to
Yuke Li (email: yklee@hznu.edu.cn).

%FIGURE 2%%%%%%%%%%%%%%%%%%%%%%%%%%%%%%%%%%%%%%%%%%%%%%%%%%%%%%%%%%%%%%%%%%%%%%%%%%%%%%%%%%%%%%%%%%%%%%%%%%%%%%%%%%%%%%%%%%%%%%%%%%%%%%%%%%%%%%%%%%%%%%

%%
\end{document}